\documentclass[11pt,twoside]{article}
\usepackage{asp2006}
\usepackage{psfig}
\usepackage{epsf}
\usepackage{graphics}
\usepackage{lscape}
\begin{document}
\title{Galaxies at High Redshift and Reionization}

\author{ Andrew Bunker$^{1,2,3}$, Elizabeth Stanway$^4$, Richard Ellis$^5$, Mark Lacy$^5$, 
Richard McMahon$^6$, Laurence Eyles$^1$, Daniel Stark$^{4,6}$,
Kuenley Chiu$^{1,2,6}$}

\affil{$^1$University of Exeter, School of Physics, Stocker Road, Exeter, 
EX4\,4QL, U.K.\\ 
$^2$ Anglo-Australian Observatory, P.O.\ Box 296, Epping, NSW 1710, Australia\\
$^3$ Department of Physics, Denys Wilkinson Building, Oxford, OX1\,3RH, U.K.\\
$^4$H.~H.\ Wills Physics Laboratory, Tyndall Ave., Bristol, BS8\,1TL, U.K.\\
$^5$California Institute of Technology, Pasadena, CA 91125, U.S.A.\\
$^6$Institute of Astronomy, Madingley Road, Cambridge CB3 0HA, 
U.K.\\
E-mail: a.bunker1@physics.ox.ac.uk, E.R.Stanway@bristol.ac.uk, rse@astro.caltech.edu,  mlacy@ipac.caltech.edu, 
rgm@ast.cam.ac.uk, eyles@astro.ex.ac.uk,dps@ast.cam.ac.uk, chiu@astro.caltech.edu}

\begin{abstract}
The quest to discover the most distant galaxies has developed rapidly in the last decade. We are now exploring redshifts of 6 and beyond, when the Universe was less than a billion years old, an epoch when the previously-neutral intergalactic medium was reionized.
The continuing discovery of galaxies at progressively higher and higher redshifts has been driven by the availability of large telescopes on the ground and in space, improvements in detector technology, and new search strategies. Over the past 4 years, the Lyman break technique has been shown
to be effective in isolating $z\approx 6$ star-forming $i'$-drop galaxies through 
spectroscopic confirmation with large ground-based telescopes
(Keck, Gemini and the ESO VLTs). Narrow-band imaging, notably with the wide field of the Subaru telescope,
 has also produced samples of Lyman-$\alpha$
emitters at these redshifts. Analysis of the {\em Hubble Ultra
  Deep Field} (HUDF -- the deepest images obtained so far, and likely to
  remain so until the {\em James Webb Space Telescope, JWST}), has enabled us to explore
the faint end of the luminosity function, which may contribute the bulk
of the total star formation. The discovery of this $i'$-drop galaxy
population has been used to infer the global star formation rate density at
this epoch ($z\approx 6$), and we are now begining to constrain the contribution to
reionization of the UV flux from these galaxies. Infrared
data from the {\em Spitzer Space Telescope} has been used to determine the spectral
energy distributions (SEDs) from the rest-frame UV to the optical of some $i'$-drops, and constrain
the previous star formation histories, masses and age of these sources. 
The indications are that much of the stellar mass of these galaxies might
have formed in vigorous bursts at $z>6$. The
next big advances would be to test the population synthesis modelling of these
$z\sim 6$ galaxies
through spectroscopy of the rest-frame optical (rather than crude broad-band SEDs),
and also to push the observational horizon for galaxies further to directly explore
star formation during the reionization epoch. {\em JWST} is likely to have
a profound impact on realising these goals.
\end{abstract}

\section{Introduction}

There has been enormous progress over the past decade in discovering
galaxies and QSOs at increasingly high redshifts. We are now probing
far enough back in time that the Universe at these early epochs was
fundamentally different from its predominantly ionized state today.
Observations of $z>6.2$ QSOs (Becker et al.\ 2001, Fan et al.\ 2002)
show near-complete absorption of flux at wavelengths short-ward of
Lyman-$\alpha$ (Gunn \& Peterson 1965), indicating that the Universe
is optically thick to this line, and that the neutral fraction of
hydrogen is much greater than at lower redshifts. {\em WMAP} results from the 
cosmic microwave background indicate
that the Universe was completely neutral at redshifts of $z\sim 10$
(Spergel et al.\ 2007).  There is an ongoing debate as to what reionized
the Universe at $z>6$: is it AGN or ionizing photons (produced in hot,
massive, short-lived stars) escaping from star forming galaxies? AGN appear
to be very under-abundant at these epochs (e.g., Dijkstra, Haiman \&
Loeb 2004), so to address the reionization issue it is crucial to
know the global star formation rate at high redshifts, along
with the escape fraction for these ionizing photons. 
In this review talk I will outline the work done
in identifying star-forming galaxies within
the first billion years of the Big Bang through deep imaging in the optical and near-infrared with the
{\em Hubble Space Telescope (HST)} and spectroscopy with large telescopes from the ground. I will also discuss what can be
learned about the stellar masses and formation epochs of these
galaxies, and what role they may have played in the reionization
of the Universe.
The standard ``concordance'' cosmology is used throughout
($\Omega_M=0.3$,
$\Omega_{\Lambda}=0.7$, and use $h_{70}=H_0/70\,{\rm
km\,s^{-1}\,Mpc^{-1}}$). All magnitudes are on the $AB$ system.

\section{The Lyman-Break Technique at $z\approx 6$}

To address the contribution of star-forming galaxies to reionization,
we need to reliably identify galaxies at $z>5$ and measure their
integrated rest-frame UV flux. Magnitude-limited spectroscopic surveys have shown that the high-redshift tail is small, so simply taking spectra of everything in a deep field is an expensive way of discovering a handful of high-redshift galaxies. Some way of isolating candidate high-redshift galaxies from the foreground population is required to provide targets for follow-up spectroscopic confirmation.
One technique is to use narrow-band imaging to identify galaxies where highly-redshifted line emission falls within the passband. After many years of fruitless searches in the 1980s and early 1990s, this method is now yielding great results, although the Lyman-$\alpha$ line typically targetted is a poor measure of the star formation rate. In the past decade, deep imaging in several different broad-bands (typically spanning the optical and near-infrared) has also been used to estimate the redshifts of galaxies on the basis of their observed colours -- essentially using the few photometric datapoints at different wavelengths as a crude low-dispersion spectrum. In this article I will focus on a particular variant of these ``photometric redshifts", which relies on a strong spectral break arising from absorption by neutral hydrogen in intervening clouds along the line of sight (the Lyman-$\alpha$ forest). This ``Lyman break
technique'' was originally pioneered at $z\sim 3$ 
by Steidel and co-workers (Steidel, Pettini \& Hamilton 1995; Steidel et
al.\ 1996) who identified $U$-band drop-outs.
At $z\sim 3$
the technique involves the use of three filters: the $U$-band below the Lyman
limit ($\rm \lambda_{rest}=912$\,\AA ); the $g$- or $B$-band in the Lyman forest region
and a third filter, the $R$-band, long-ward of the Lyman-$\alpha$ line ($\rm
\lambda_{rest}=1216$\,\AA). 
At $z\approx 6$, we can efficiently use only two filters, above
($z'$-band 9000\AA) and below ($i'$-band 8000\AA ) the continuum break
at the Lyman-$\alpha$ line ($\rm\lambda =(1+z)\times1216$\,\AA ); since the
integrated optical depth of the Lyman-$\alpha$ forest at $z\approx 6$ is $\gg 1$ (see
Figure~1), there is essentially no flux at shorter
wavelengths, rendering the shortest-wavelength filter below the Lyman limit
redundant.  The key issue is to work at a sufficiently-high
signal-to-noise ratio that Lyman break ``drop-out'' galaxies can be
safely identified through detection in a single redder band (i.e.,
$z'$-band). 
This approach has been demonstrated to be effective by the
SDSS collaboration in the detection of $z\approx 6$ QSOs using the
$i'$- and $z'$-bands alone (Fan et al.\ 2002), using wide-area but
shallow ground-based imaging, which is sensitive to bright but rare quasars.
Several deep imaging programmes with the Advanced Camera for Surveys (ACS)
on {\em HST} use the sharp-sided SDSS F775W ($i'$) and F850LP
($z'$) filters, which can be used to locate the``$i$-drop''  candidate
$z\approx 6$ galaxies (fainter but more numerous than the QSOs).
Figures~1\,\&\,2 illustrates how a
colour cut of $(i'-z')_{AB}>1.3$ can be effective in selecting sources
with $z>5.6$. The ACS images of the {\em Chandra} Deep Field South
from the {\em HST} Treasury ``Great Observatory Origins Deep Survey''
(GOODS; Giavalisco \& Dickinson 2003) have been used to discover
this $i'$-drop population of $z\approx 6$ galaxies (Stanway, Bunker \& McMahon
2003; Dickinson {\em et al.\ }2004; Giavalisco {\em et al.\ }2004).
Stanway, Bunker \& McMahon (2003) searched 150\,arcmin$^{2}$ for
$(i'-z')_{AB}>1.5$ $i'$-drop objects, of which six were probable $z>5.7$
galaxies brighter than $z_{AB}<25.6$.  
To address potential cosmic variance issues, a similar analysis in the
GOODS-North field, which yielded a consistent estimate of the surface
density of $z\simeq 6$ star forming sources (Stanway et al.\ 2004a).
Since these initial results, larger numbers of $i'$-drops have been
uncovered in additional {\em HST}/ACS pointings (e.g., Bouwens
et al.\ 2006) and from the ground (e.g., the Subaru Suprime-Cam --
Iwata et al.\ 2007) as well as deeper imaging of the GOODS fields,
particularly the {\em Hubble Ultra Deep Field} (Bunker et al.\ 2004) within GOODS-South.

\begin{figure}
\resizebox{0.49\textwidth}{!}{\includegraphics{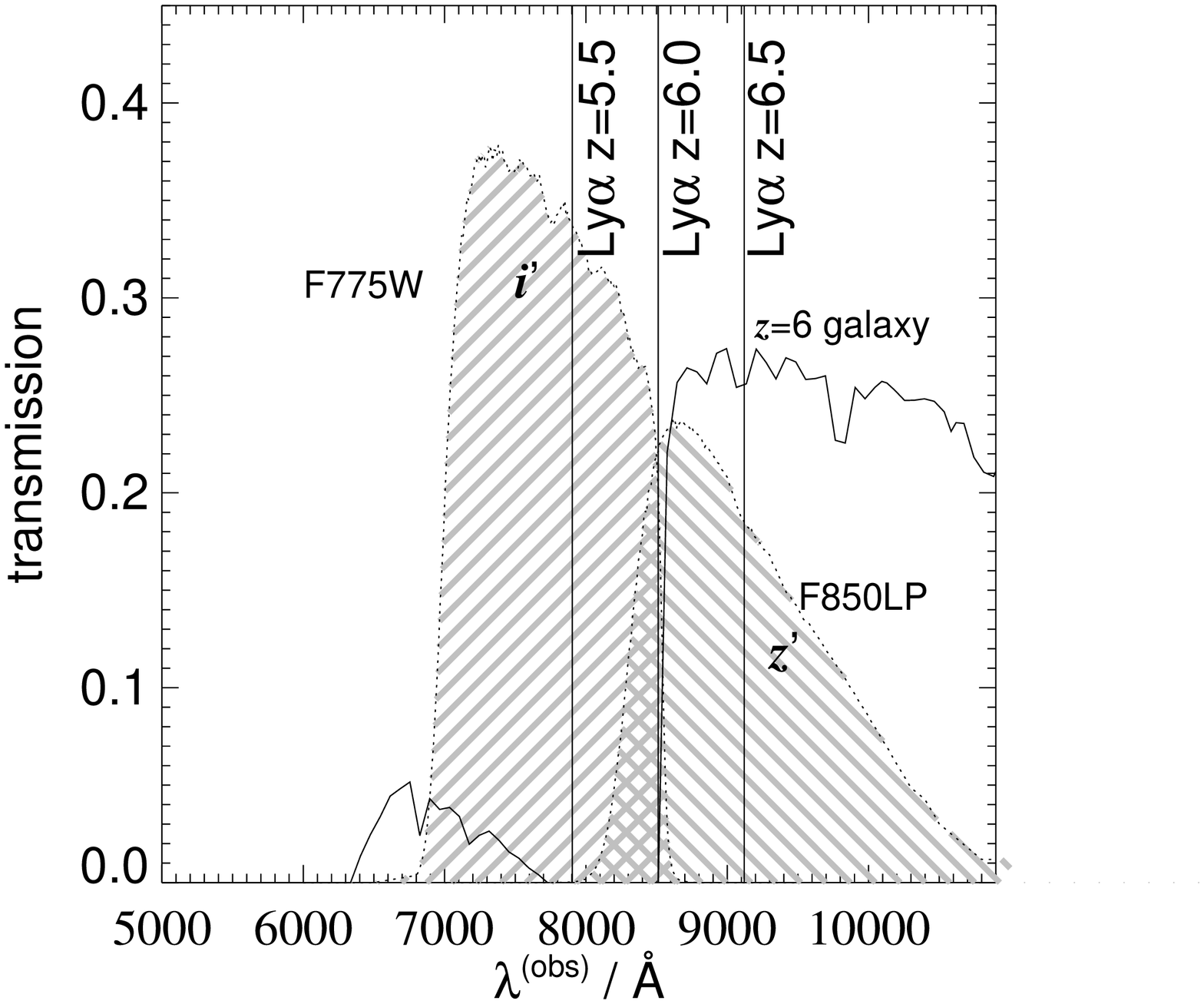}}
\resizebox{0.49\textwidth}{!}{\includegraphics{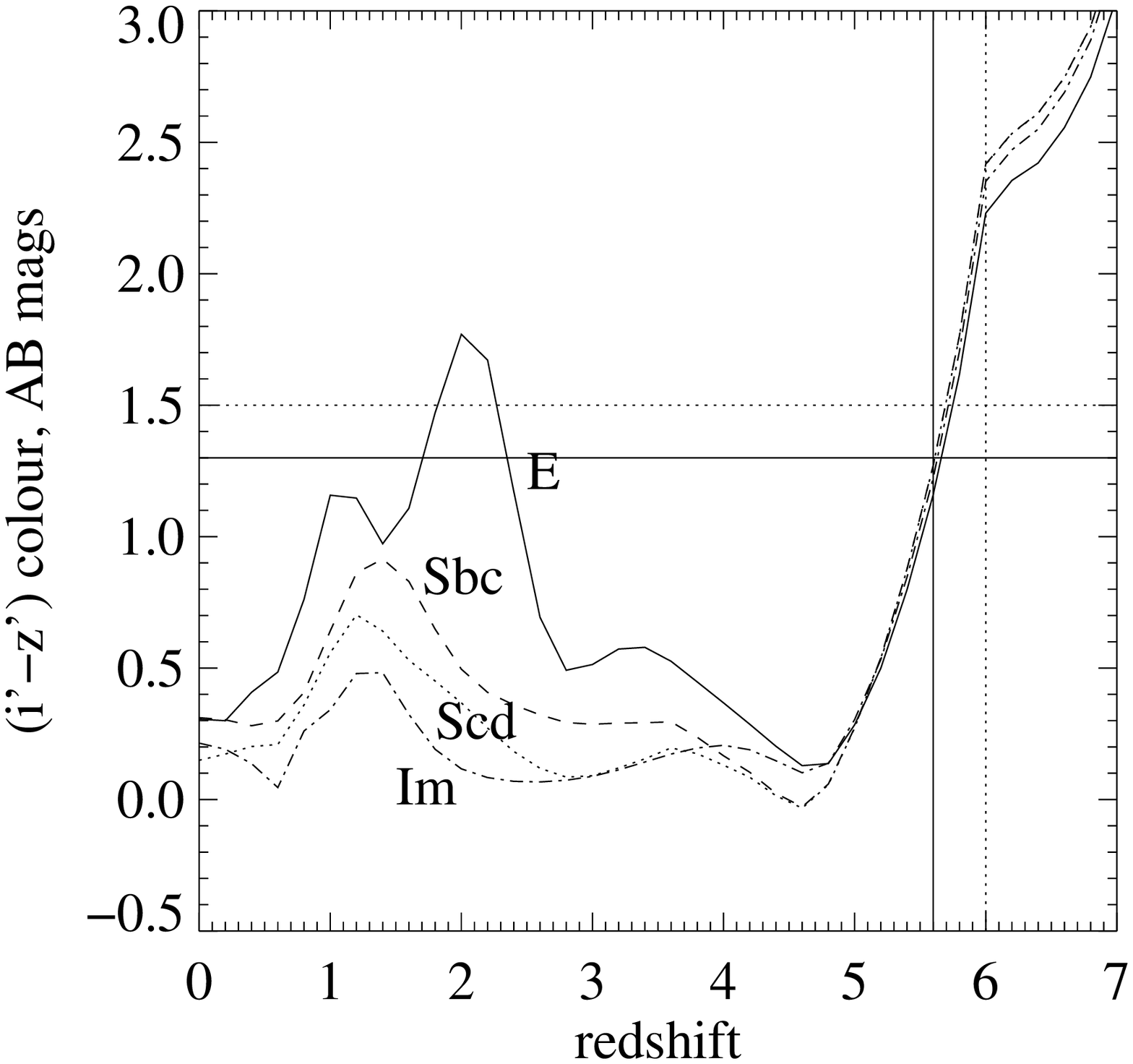}}
\caption{{\bf Left:} The ACS-$i'$ and -$z'$ bandpasses overplotted on
the spectrum of a generic $z=6$ galaxy (solid line), illustrating the
utility of our two-filter technique for locating $z\approx 6$
sources. {\bf Right:} Model colour-redshift tracks for galaxies with
non-evolving stellar populations. 
(from the template spectra of Coleman, Wu \& Weedman 1980). 
The contaminating `hump' in the
$(i'-z')$ colour at $z\approx 1-2$ arises when the Balmer break and/or
the 4000\,\AA\ break redshifts beyond the $i'$-filter.}
\label{fig:filters}
\label{fig:tracks}
\end{figure}

\section{Spectroscopic Confirmation of $z\approx 6$ Galaxies}

The effectiveness of the Lyman-break technique has been demonstrated
at lower redshifts through spectroscopic confirmation of hunderds of
$U$-band dropouts at $z\approx 3$ and $B$-band drop-outs at $z\approx
4$ (Steidel et al.\ 1999).  However, it is important to demonstrate that the $i'$-drop
selection is similarly isolating galaxies at $z\approx 6$, if we are
to use the surface density of $i'$-drops to draw global inferences
about galaxies within the first billion years. A known possible
contaminant is the Extremely Red Object (ERO) population of evolved
galaxies at $z\approx 1-2$ which can produce large $(i'-z')$ colours:
deep near-infrared imaging should identify EROs, and this 
possible contamination is considered in Stanway, McMahon \& Bunker (2005)
from the {\em HST}/NICMOS survey of the HUDF. Low-mass
Galactic stars (M/L/T dwarfs) are another interloper population.
Deep spectroscopy with Keck/DEIMOS of the
brightest $i'$-drop identified by Stanway, Bunker \& McMahon (2003)
in GOODS-South revealed the brightest one, galaxy SBM03\#3 ($z'=24.7$\,mag), 
to be at $z=5.78$ (Bunker et al.\ 2003; Dickinson et al.\ 2004) and
another $i'$-drop (SBM03\#1, $z'=25.4$) was found to lie at a similar
redshift of $z=5.83$ (Stanway et al.\ 2004ab; Dickinson et al.\ 2004).
The $i'$-drop spectra typically show a single emission line, with
no significant continuum detected at moderate spectral resolution ($R\sim 1000-5000$); 
spectrally-resolved profiles of the
emission lines are asymmetric (as high-$z$ Lyman-$\alpha$ tends to be)
with a P-Cygni-like profile and a sharp cut-off on the blue wing (Figure~2b).  
The line fluxes of these brighter $i'$-drops are $\approx 10^{-17}\,{\rm ergs\,cm}^{-2}\,{\rm
s}^{-1}$. The equivalent widths are $W_{\rm rest}=20-30$\,\AA\ using
the $z'$-band photometry from HST/ACS, which is within the range seen
in high-$z$ star-forming galaxies at $z\sim 3-4$. The velocity widths of the
Lyman-$\alpha$ are comparatively narrow ($v_{\rm FWHM}=300\,{\rm
km\,s}^{-1}$) and we do not detect the high-ionization line
N{\scriptsize~V}\,$\lambda$\,1240\,\AA , which strongly support the
view that this line emission is powered by star formation rather than
an AGN. 
Subsequent spectroscopic 
programmes, including using FORS\,2 on the ESO VLT (Vanzella et al.\ 2006) and GMOS ``nod\,\&\,shuffle'' 
spectroscopy on Gemini
(GLARE, Stanway et al.\ 2004b, 2007) have confirmed many more $z\approx 6$ galaxies in
GOODS-South. Slitless spectroscopy of the HUDF with {\em HST}/ACS
(GRAPES, Malhotra et al.\ 2005) has also confirmed the spectral breaks
of the $i'$-drop galaxies, although the spectral resolution of
$R\approx 70$ is usually insufficient to see the Lyman-$\alpha$ emission
lines in all but the most extreme equivalent widths.

Although such spectroscopy has proven that the $i'$-drop selection technique does select star-forming
galaxies at $z\approx 6$, the efforts to date have been limited despite pushing the limits
of 10-m class telescopes from the ground. Only a fraction of the $\sim 500$ $i'$-drops imaged
with {\em HST} (see the compilation of Bouwens et al.\ 2006) have spectroscopic redshifts.
It seems likely that Lyman-$\alpha$ line emission does not emerge from some galaxies,
through resonant scattering and absorption by dust (a well-known result at lower redshifts,
e.g., Steidel et al.\ 1999), although the fraction of galaxies at $z\sim 6$ with large Lyman-$\alpha$ equivalent widths seems greater than at $z\sim 3$ (Stanway et al.\ 2007), possibly implying low-metallicity, a top-heavy initial mass function, or extremely young ages for the starbursts.

With its multi-object spectroscopic capability using micro-shutter technology, NIRSpec on {\em JWST} (Gardner et al.\ 2006) will have the capability to determine
redshifts for most of the $i'$-drop $z\sim 6$ candidates identified, even to the depth of the HUDF. 
Beyond simply extending ground-based
Lyman-$\alpha$ spectroscopy at $z\sim 6$ to fainter galaxies, NIRSpec will 
have the sensitivity to detect the rest-frame UV and optical continuum. Studies of
interstellar medium (ISM) and stellar photospheric lines of $U$-drop galaxies
at $z\sim 3$ in the rest-UV have provided clues to the initial mass function, metallicity
and outflow velocities in these galaxies (e.g., Shapley et al.\ 2003).
Only in the case of the brightest $i'$-drop have such ISM metal absorption lines
been seen (Dow-Hygelund et al.\ 2005), but NIRSpec will enable the whole population
to be studied with its intermediate-resolution and high-resolution modes ($R\approx 1000$
and $R\approx 3000$). Perhaps most importantly, NIRSpec will not be restricted to the rest-UV
at $z\sim 6$, which is the case with ground-based optical/near-IR spectroscopy.
NIRSpec has a coverage from $0.8-5\,\mu$m, unhampered by the OH sky lines and large
thermal background seen from ground-based observatories. Hence, for the $z\sim 6$ population,
we can target the full range of emission lines from Lyman-$\alpha$\,1216\,\AA\ to 
H$\alpha$\,6563\,\AA . Emission line diagnostics can then be used to estimate the 
metallicities and the extinction due to dust.

\begin{figure}
\resizebox{0.49\textwidth}{!}{\includegraphics{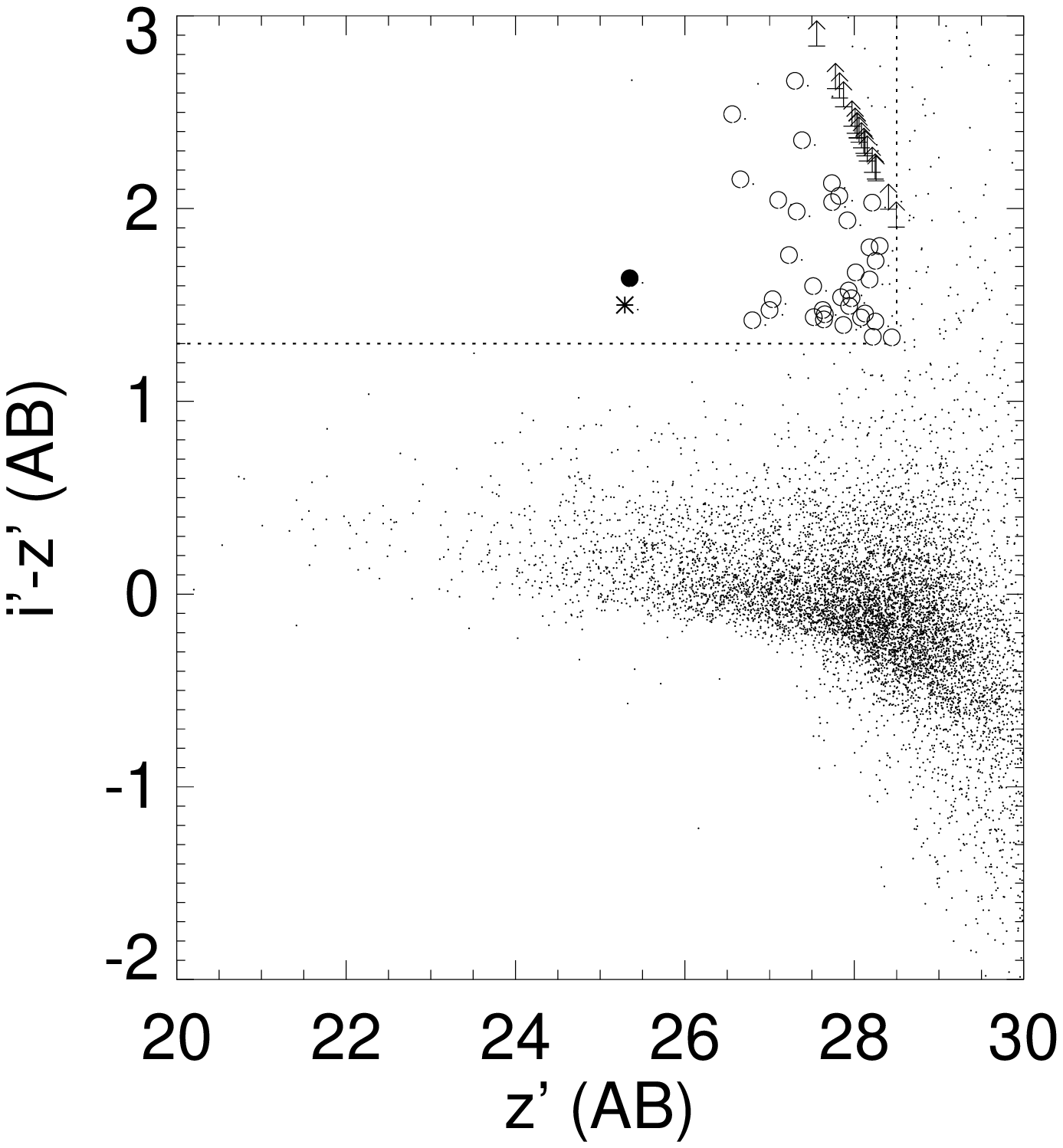}}
\resizebox{0.49\textwidth}{!}{\includegraphics{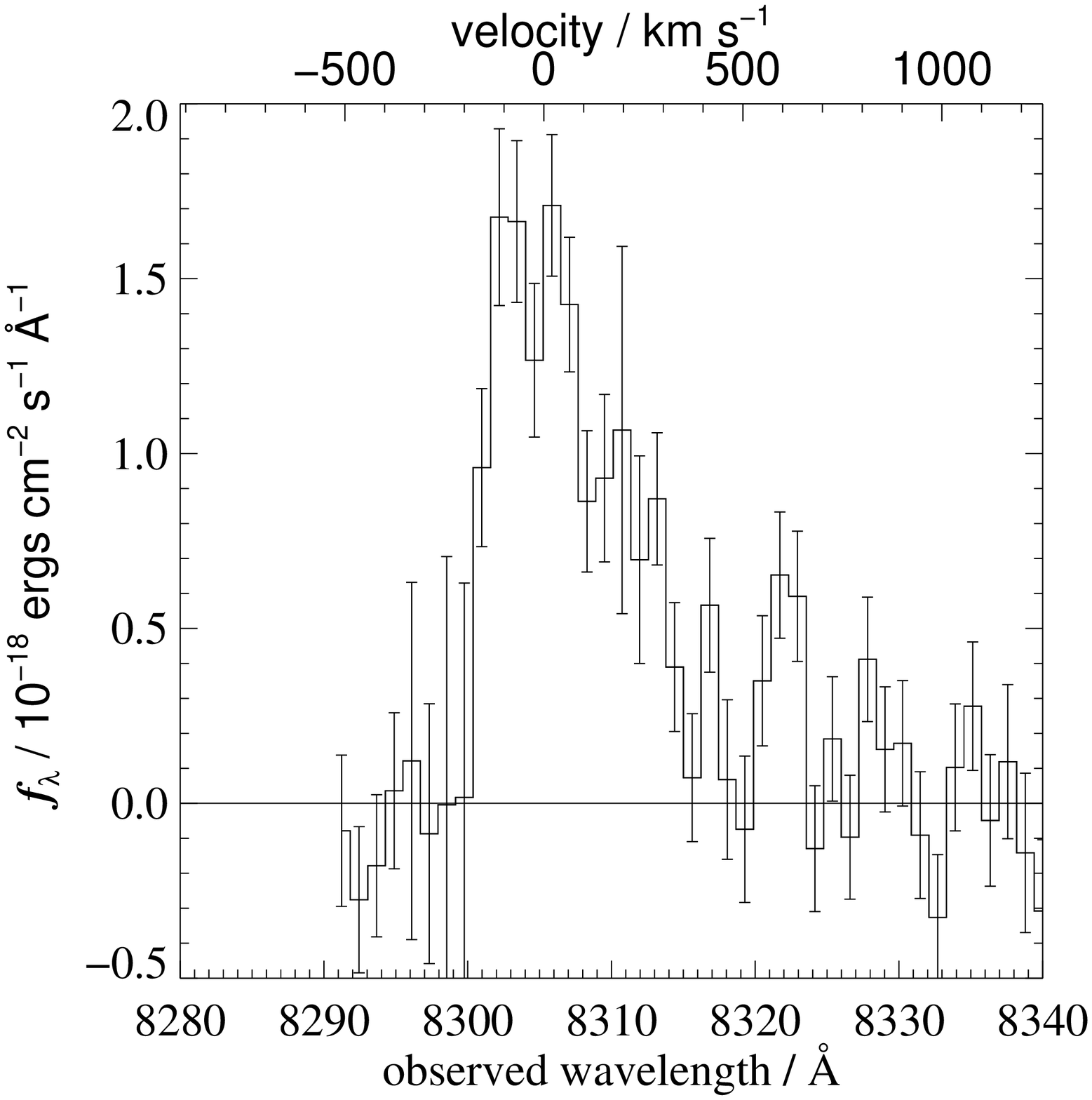}}
\caption{{\bf Left:} Colour-magnitude diagram for the HUDF data with the
limit $z'_{AB}<28.5$ and $(i'-z')_{AB}=1.3$ colour cut shown (dashed
lines). Such a catalogue could be contaminated by cool stars, EROs and
wrongly identified extended objects and diffraction spikes but
nonetheless provides a secure upper limit to the abundance of z$\approx
6$ star forming galaxies.  Circles and arrows (lower limits) indicate
our $i'$-drop candidate $z\approx 6$ galaxies. The asterisk is the only
unresolved $i'$-drop in our HUDF sample, a probable star. The solid
circle is the brightest $i'$-drop in the HUDF, the
spectroscopically-confirmed galaxy SBM03\#1 at $z=5.83$, with our
discovery spectrum from Keck/DEIMOS shown ({\bf right}, Stanway et al.\
2004a, also confirmed by Dickinson et al.\ 2004).}
\label{fig:izzcolmag}
\resizebox{0.49\columnwidth}{!}{\includegraphics{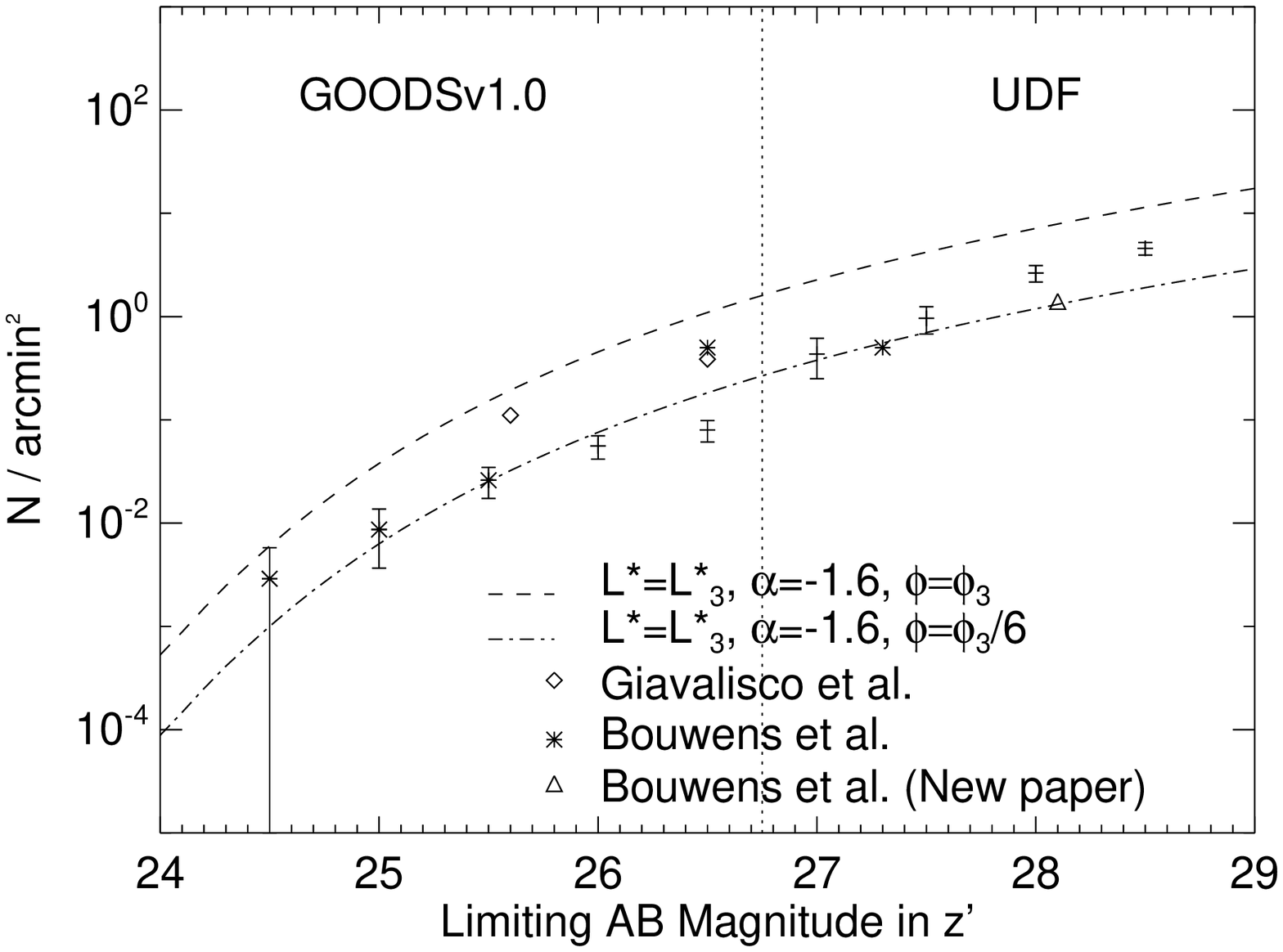}}
\resizebox{0.49\textwidth}{!}{\includegraphics{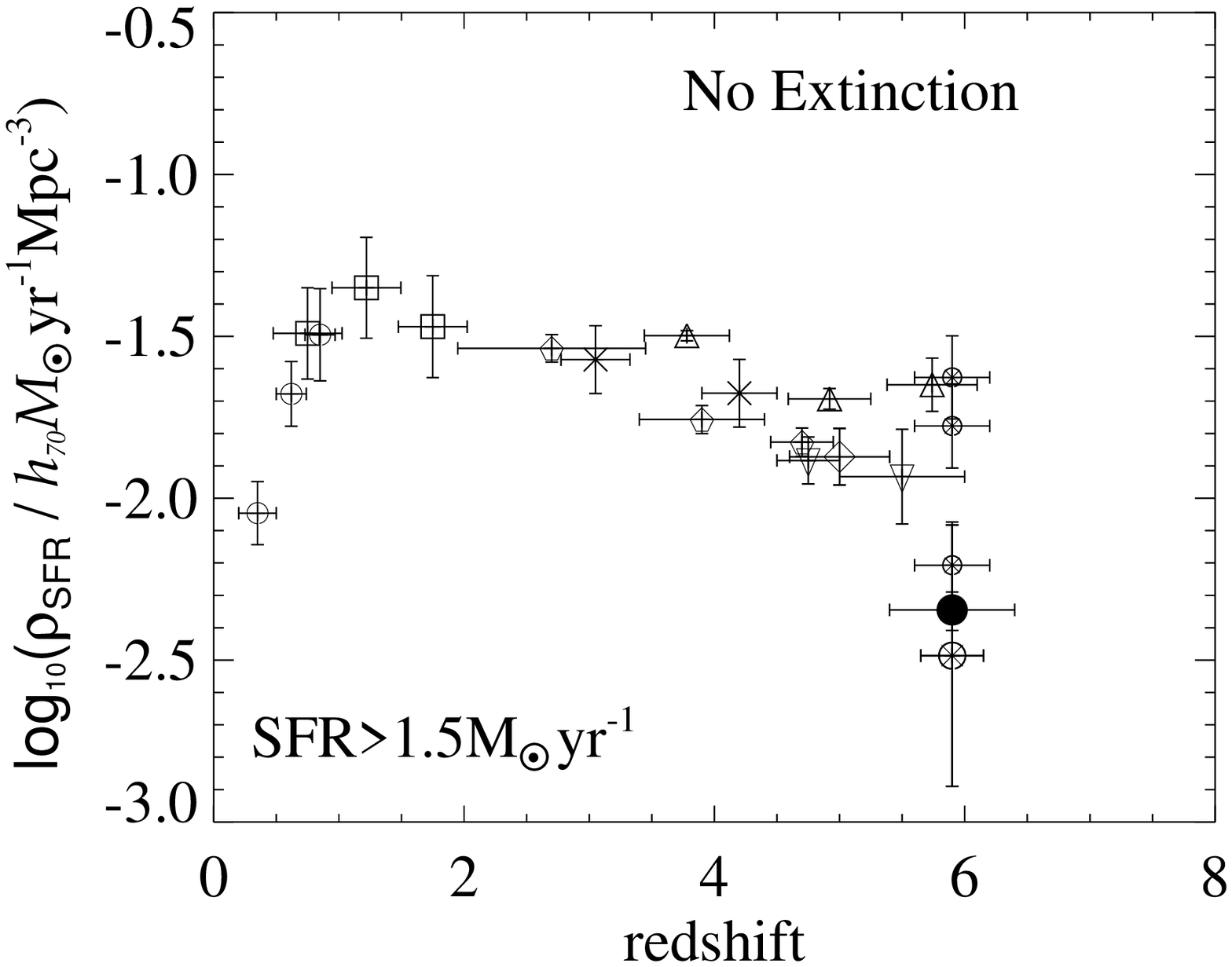}}
\caption{{\bf Left:} Cumulative source counts per arcmin$^{2}$ of $i'$-dropouts
as a function of $z'$-band magnitude. The HUDF data (area of
11\,arcmin$^{2}$ for $z'_{AB}\ge 27.0$) is compared
with our $z'_{AB}\le 25.6$ single epoch GOODSv0.5 ACS/WFC imaging over
300\,arcmin$^{2}$ (Stanway, Bunker \& McMahon 2003), and combined 5 epoch GOODSv1.0
images to $z'_{AB}<27.0$ (Stanway 2004, PhD.\ thesis, Cambridge). Also plotted are the surface densities of $i'$-drops from
Giavalisco et al.\ (2004)
Bouwens et al.\ (2003, 2004)
\newline
{\bf Right:} An updated version of the `Madau-Lilly' diagram (Madau et al.\
1996; Lilly et al.\ 1996) illustrating the evolution of the comoving
volume-averaged star formation rate.  Our work from the HUDF data is
plotted a solid symbol. Other determinations have been recalculated for
our cosmology and limiting UV luminosity of
$1.5\,h_{70}^{-2}\,M_{\odot}\,{\rm yr}^{-1}$ at $z=6$. 
See Bunker et al.\ (2004) for definition of the other symbols.
}
\label{fig:numcounts}
\label{fig:madau_plot}
\end{figure}

\section{The star formation rate density at $z\approx 6$}
\label{sec:pol}

Estimates of the star formation rate density at the highest
redshifts have relied on the intensity of
the rest-frame UV continuum emission, redshifted into the $z'$-band at
$z\approx 6$; this is dominated by the light from the most massive
stars ($M>10\,M_{\odot}$ which are the hottest and bluest stars),
which are the shortest lived, and hence a tracer of the instantaneous
star formation rate. To go from the number of massive stars formed to
the total star formation rate requires an assumption about the stellar
initial mass function (IMF). In the absence of dust obscuration, the
relation between the flux density in the rest-UV around $\approx
1500$\,\AA\ and the star formation rate (${\rm SFR}$ in
$M_{\odot}\,{\rm yr}^{-1}$) is given by $L_{\rm UV}=8\times 10^{27}
{\rm SFR}\,{\rm ergs\,s^{-1}\,Hz^{-1}}$ from Madau, Pozzetti \&
Dickinson (1998) for a Salpeter (1955) IMF with
$0.1\,M_{\odot}<M^{*}<125\,M_{\odot}$. The relatively bright magnitude
cut of $z'_{AB}<25.6$ for the original $i'$-drop selection from
the GOODSv0.5 data (Stanway, Bunker \&
McMahon 2003; Stanway et al.\ 2004a) corresponds to an unobscured star
formation rate of $15\,h^{-2}_{70}\,M_{\odot}\,{\rm yr}^{-1}$ at
$z=5.9$ (the luminosity-weighted average redshift), equivalent to
$L^{*}_{UV}$ for the $U$-band dropout population at $z\approx 3$.

A tantalizing result from our work so far is that at $z\sim 6$ there
are far fewer UV-luminous star forming galaxies than would have been
predicted if there was no evolution, based on a comparison to the
well-studied $z\sim 3-4$ Lyman break population (Steidel et al.\
1999).  In fact, the volume averaged (comoving) star formation rate in
galaxies with $>15\,h^{-2}_{70}\,M_{\odot}\,{\rm yr}^{-1}$
is $\approx 6\times$ {\em less} at $z\approx 6$ than at $z\approx 3$. So
it would appear that at this crucial epoch, where abundant sources of
UV photons are needed to reionize the Universe, there is in fact a
deficit of luminous star forming galaxies.  The bright end of luminosity function has evolved greatly from
$z\sim 6$ to $z\sim 3$. Other groups have claimed less dramatic
evolution or even no evolution in the total volume-averaged star formation
rate, based on the same fields (Giavalisco et al.\ 2004; Dickinson et
al.\ 2004) and similar {\em HST}/ACS data sets (Bouwens et al.\ 2003;
Yan, Windhorst \& Cohen 2003), but these groups work closer to the detection
limit of the images and introduce large completeness corrections for
the faint source counts. To do a complete inventory of the UV light
from star formation we must 
address the contribution of low-luminosity star-forming galaxies to
the ionizing flux.  The public availability of the {\em Hubble Ultra Deep
Field} (HUDF; Beckwith, Somerville \& Stiavelli) can address this puzzle
by pushing down the luminosity function at $z\approx 6$ to well below
the equivalent of $L^{*}$ for the $z\approx 3$ population.

\section{The Hubble Ultra Deep Field}

The {\em Hubble Ultra Deep Field} (HUDF) was a Cycle 12 STScI Director's
 Discretionary Time programme executed over September 2003 -- January
 2004, comprising 400 orbits in 4 broad-band filters (including F775W
 $i'$-band for 144 orbits; F850LP $z'$-band for 144 orbits). As the HUDF
 represents the deepest set of images yet taken, significantly deeper
 than the $I$-band exposures of the {\em Hubble Deep Fields} (Williams et
 al.\ 1996), and adds the longer-wavelength $z'$-band, it is uniquely
 suited to the goals of our program.

We performed the first analysis of $i'$-drops in the UDF (Bunker et
al.\ 2004), presenting details of 54 candidate star forming galaxies
at $z\approx 6$ in a preprint the day after the release of the UDF
data. Our analysis was subsequently independently repeated by Yan \&
Windhorst (2004), with agreement at the 98\% level (see Bunker \&
Stanway 2004 for a comparison). We take our magnitude limit as
$z'_{AB}<28.5$ (a $10\,\sigma$ cut).We measure a star formation
density of $0.005\,h_{70}\,M_{\odot}\,{\rm yr}^{-1}\,{\rm Mpc}^{-3}$
at $z\approx 6$ from galaxies in the UDF with
SFRs\,$>1.5\,h_{70}^{-2}\,M_{\odot}\,{\rm yr}^{-1}$ (equivalent to
$0.1\,L^{*}_{UV}$ at $z\approx 3$). This confirmed the previous discovery of a relatively low co-moving star formation rate density at $z\approx 6$ compared to the more recent past at $z\sim 3$ (Figure~3).

At the relatively bright cut of $z'_{AB}<25.6$ used in Stanway, Bunker
\& McMahon (2003) from the GOODSv0.5 survey, the UDF data is 98\%
complete for sources as extended as
$r_{h}=0.5$\,arcsec. Interestingly, we detect no extended (low surface
brightness) $i'$-drops to this magnitude limit in addition to the
targetted $i'$-drop SBM03\#1 in the deeper UDF data. This supports our
assertion that the $i'$-drop population is predominantly compact and
there cannot be a large completeness correction arising from extended
objects (c.f.\ Lanzetta et al. 2002). The ACS imaging is of course
picking out H{\scriptsize II} star forming regions, and these
UV-bright knots of star formation are typically $<1$\,kpc
($<0.2$\,arcsec at $z\approx 6$) even within large galaxies at low
redshift.

It is interesting that the level of stellar contamination in the UDF
$i'$-drops is only 2\%, compared with about one in three at the bright
end ($z'_{AB}<25.6$; Stanway, Bunker \& McMahon 2003; Stanway et al.\
2004). This may be because we are seeing through the Galactic disk at
these faint limiting magnitudes to a regime where there are no stars
(see also Pirzkal et al.\ 2005).

\section{Implications for Reionization}

We compare our $i'$-drop luminosity function with the work of Madau,
Haardt \& Rees (1999) for the density of star formation required for reionization.
We have updated their equation 27 for the more recent
concordance cosmology estimate of the baryon density of Spergel et al.\ (2007), and for the
predicted mean redshift of our sample ($z=6.0$):
\begin{equation}
{\dot{\rho}}_{\rm SFR}\approx \frac{0.026\,M_{\odot}\,{\rm yr}^{-1}\,{\rm Mpc}^{-3}}{f_{\rm esc}}\,\left( \frac{1+z}{7}\right) ^{3}\,\left( \frac{\Omega_{b}\,h^2_{70}}{0.0457}\right) ^{2}\,\left( \frac{C}{30}\right)
\end{equation}

This relation is based on the same Salpeter Initial Mass Function as we
have used in deriving our volume-averaged star formation rate.  $C$ is
the concentration factor of neutral hydrogen, $C=\left< \rho^{2}_{\rm
HI}\right> \left< \rho_{\rm HI}\right> ^{-2}$. Simulations suggest
$C\approx 30$ (Gnedin \& Ostriker 1997). 
Even if we take $f_{\rm esc}=1$ (no absorption by
H{\scriptsize~I}) this estimate of the star formation density required
is a factor of $\approx 5$ higher than our measured star formation
density of $0.005\,h_{70}\,M_{\odot}\,{\rm yr}^{-1}\,{\rm Mpc}^{-3}$ at
$z\approx 6$ from galaxies in the UDF with
SFRs\,$>1.5\,h_{70}^{-2}\,M_{\odot}\,{\rm yr}^{-1}$.  For faint end
slopes of $\alpha ~-1.8\rightarrow-1.3$ galaxies with $L>0.1\,L^{*}$
account for $32-80$\% of the total luminosity, so would fall short of
the required density of Lyman continuum photons required to reionize the
Universe.  If the faint-end slope is as steep as $\alpha\approx -1.9$
then there would just be enough UV Lyman continuum photons generated in
star forming galaxies at $z\approx 6$ (assuming a Salpeter IMF), but the
required escape fraction for complete reionization would still have to
be implausibly high ($f_{esc}\approx 1$, whereas all high-$z$
measurements to date indicate $f_{esc}\ll 0.5$: Fern\'{a}nadez-Soto, Lanzetta
\& Chen 2003; Steidel, Adelberger \& Pettini 2001).
If star forming galaxies at redshifts close to $z=6$ were
responsible for the bulk of reionization, then a very different initial
mass function would be required, or the calculations of the clumping
factor of neutral gas would have to be significantly over-estimated
(see also Stiavelli, Fall \& Panagia 2004).
Alternatively another low-luminosity population (e.g., forming globular
clusters; Ricotti 2002) could be invoked to provide some of the
shortfall in ionizing photons.  
It is also plausible that the bulk of
reionization occurred at redshifts well beyond $z=6$: the WMAP
polarization data indicate  $z_{reion}\sim 10$ (Spergel et al.\ 2007), and it
is possible that the Gunn-Peterson troughs seen in $z> 6.2$ QSOs (Becker
et al.\ 2001; Fan et al.\ 2002) mark the very last period of a neutral
IGM.
If some of these $i'$-drop galaxies can be shown to harbour stellar
populations with ages of a few hundred Myr, then this pushes their
formation epoch to $z\sim 10$. Measurements of the stellar masses of
individual $z\sim 6$ galaxies can also constrain structure formation
paradigms; in a simple hierarchical model, massive galaxies assemble
at later times through merging, so it might be expected that in this
scenario the number density of massive evolved galaxies in the first
Gyr would be low.

Obtaining such observational evidence of old stars at high redshift is challenging: the initial work on $z\sim 6$ galaxies was based on optical and near-infrared imaging and explored the rest-frame ultraviolet (UV) in these galaxies, which is completely
dominated by recent or ongoing star formation.
The addition of imaging at wavelengths  $3.6-8\,\mu$m with the IRAC camera on the {\em
  Spitzer Space Telescope} extends the coverage to the rest-frame optical at $z\sim 6$, beyond the age-sensitive Balmer/4000\,\AA\ break and where the light is dominated by stars that form the bulk of the stellar mass. With the addition of this {\em Spitzer}/IRAC imaging, spectral energy
distributions from the multi-wavelength broad-band photometry can be fit
to stellar population synthesis models to constrain the 
stellar masses and ages, and hence
the preceding star formation history and the formation epochs can be explored.

In Eyles et al.\ (2005) we identified two $i'$-drops with spectroscopic redshifts of $z\approx 5.8$ which exhibited significant Balmer/4000\,\AA\ spectral
  breaks (Figure~4) , implying stellar ages of $\sim 200-700$\,Myr, and formation redshifts of
  $7\le z_f \le 18$. Large stellar masses in the range $\sim 1 -
  3\times 10^{10}\,M_{\odot}$ were inferred. We then studied the whole Lyman-break population in the GOODS-South field with IRAC/{\em Spitzer} at $z\approx 6$ (the $i'$-drops, Eyles et al.\ 2007) and $z\approx 5$ (the $v$-drops, Stark et al.\ 2007). Half of these are confused with
  foreground sources at {\em Spitzer} resolution, but from
  those with clean photometry we find that
  a surprisingly large fraction (40\%) have
 evidence for substantial Balmer/4000\,\AA\ spectral
  breaks. This indicates the presence of old underlying stellar
  populations that dominate the stellar masses. 
 Analysis of those $i'$-drops that are
  undetected by IRAC at 3.6\,$\mu$m indicates that a subset of the population are younger,
  considerably less massive systems.
We calculate that emission line contamination should not severely
  affect our photometry or derived results.  Using SED fits out to $8\,\mu$m, we
 find little evidence for substantial intrinsic dust reddening in our sources.

Using the constrained properties of our $i'$-drop sample, we were able
to calculate a value for the $z\sim 6$ stellar mass density of
$2.5\times 10^{6}M_{\odot}\,{\rm Mpc}^{-3}$ (Figure~5), correcting for those
objects eliminated from our analysis due to their untreatable
IRAC-confusion and those lacking GOODS-MUSIC photometric redshifts.  Using a somewhat uncertain correction in order to
account for the stellar mass in objects below our $z'$-band magnitude
selection limit, this value could perhaps be $5 - 8\times
10^{6}M_{\odot}\,{\rm Mpc}^{-3}$.  Any post-starburst and dust-obscured
$z\sim 6$ sources would not be found using the $i'$-drop selection
technique, and hence our $z\sim 6$ stellar mass density value is 
necessarily a lower limit, and is consistent with the estimates of Yan et al.\ (2006).
Exploring the previous star formation histories of our $i'$-drops, as inferred 
from their SED fitting, we suggest that the global star formation of these 
sources may have been substantially higher prior to the epoch of observation (Figure~5),
and the resultant UV flux at $z>7$ may have played an important role
in reionizing the Universe.

\begin{figure}
\resizebox{0.48\textwidth}{!}{\includegraphics{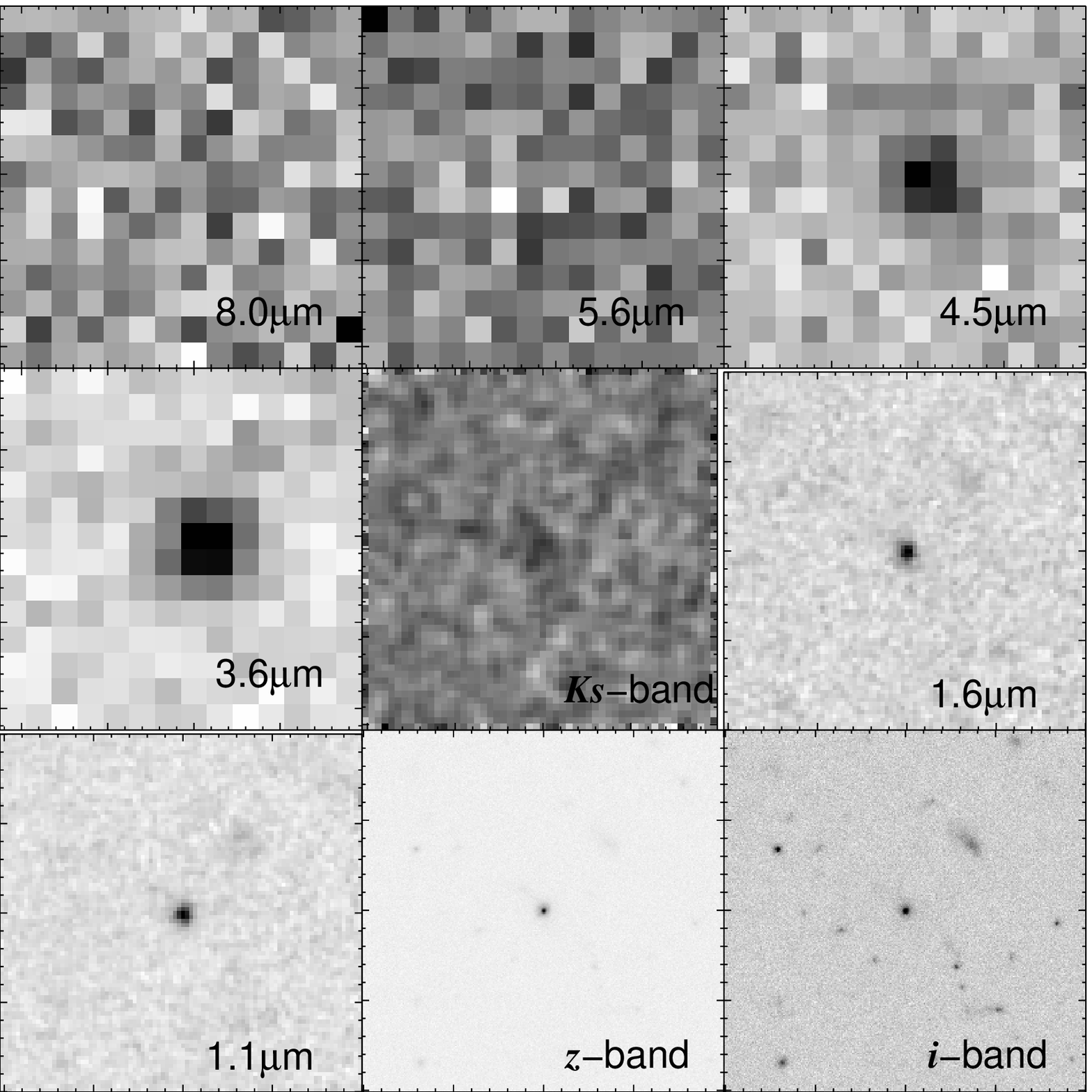}}
\resizebox{0.50\textwidth}{!}{\includegraphics{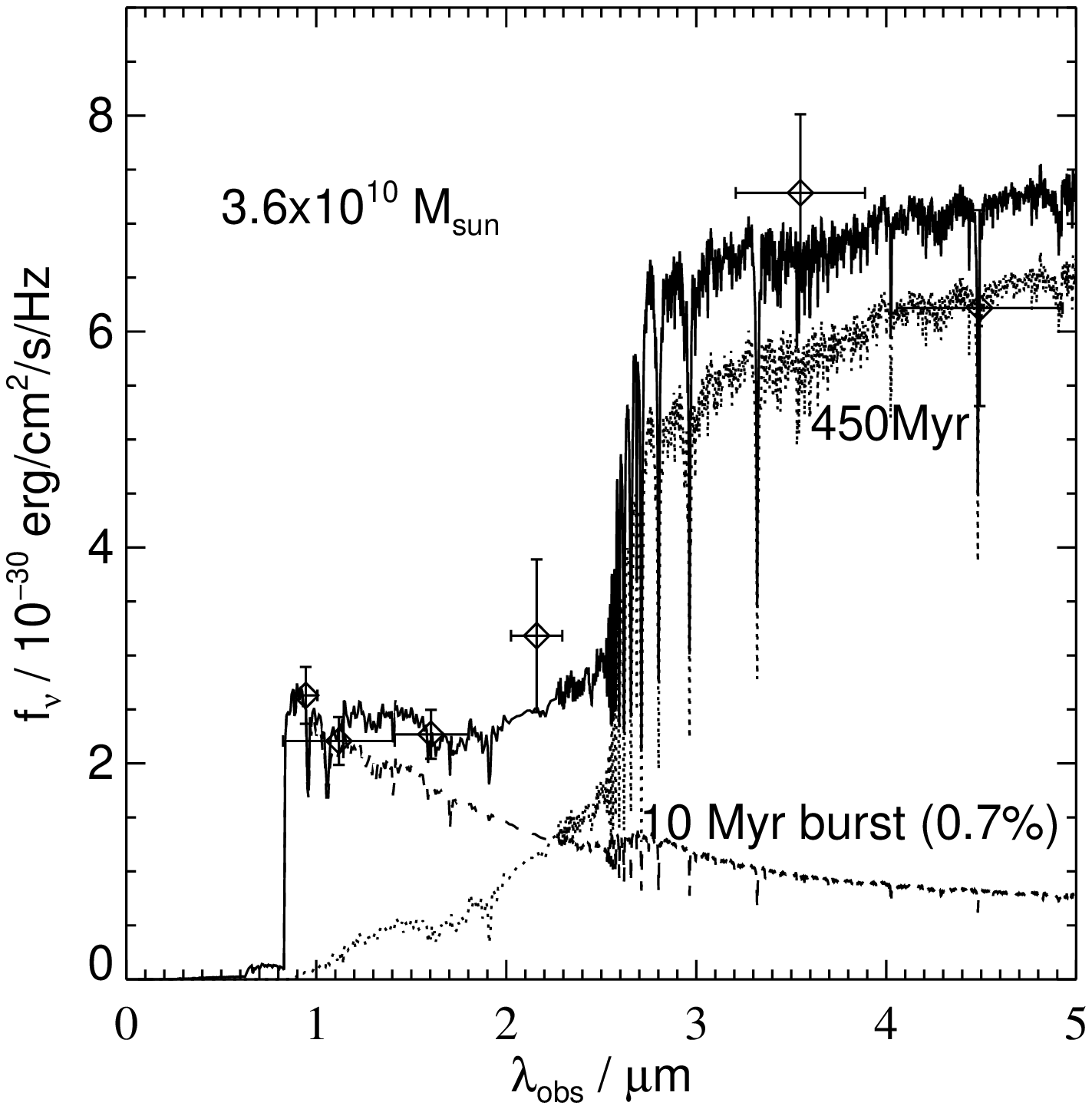}}
\caption{{\bf Left:} The {\em Spitzer} IRAC and {\em HST} ACS \&
NICMOS images of the $z=5.83$ galaxy, SBM03\#1. {\bf Right:} the {\em
Spitzer} and {\em HST} wavebands straddle the age-sensitive
Balmer/4000\,\AA\ break, and reveal an underlying old ($\sim
400$\,Myr) population which dominates the stellar mass, $\sim 3\times
10^{10}\,M_{\odot}$ (Eyles et al.\ 2005).}
\label{fig:spitzer}
\end{figure}

\begin{figure}
\resizebox{0.48\textwidth}{!}{\includegraphics{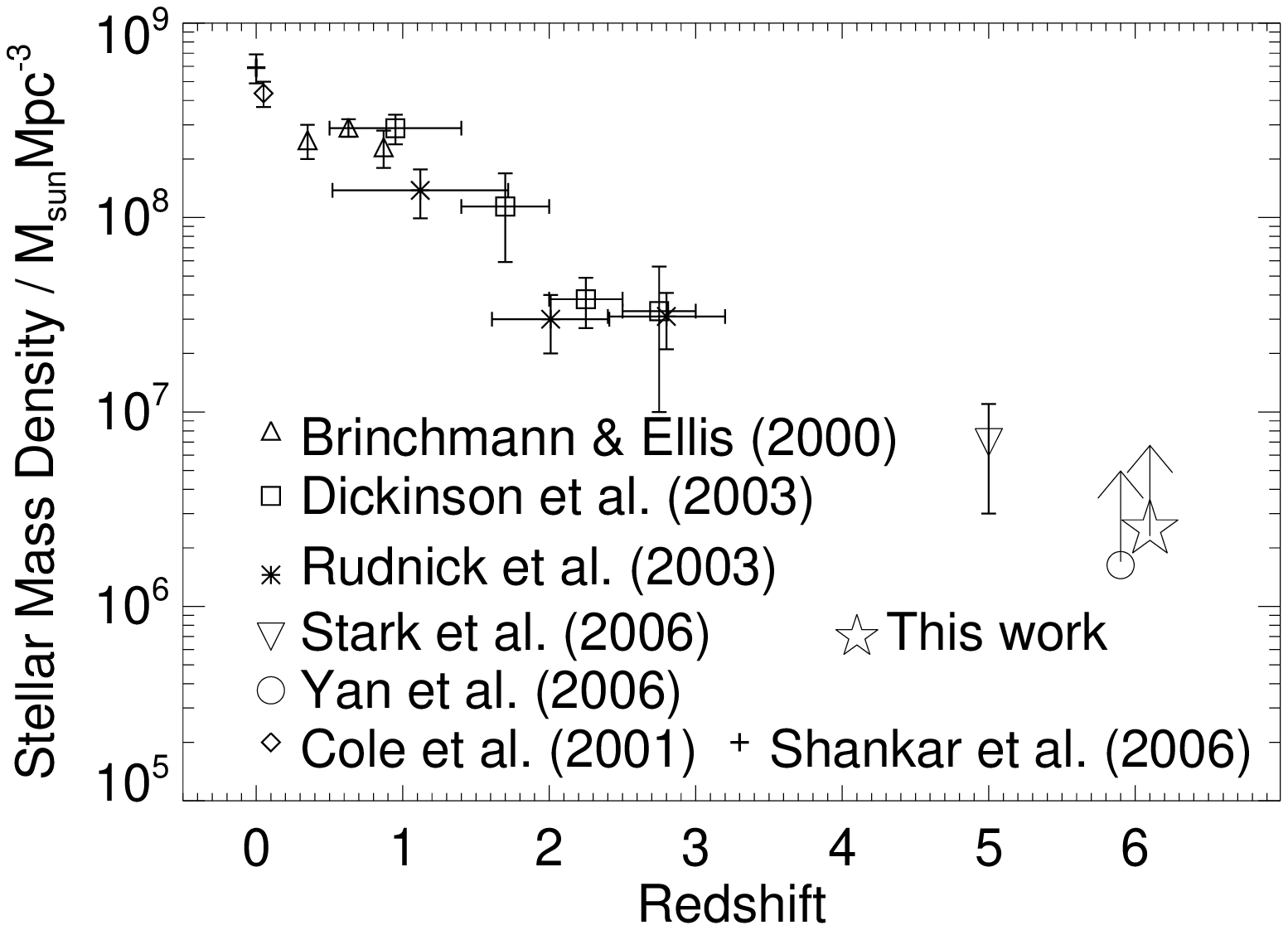}}
\resizebox{0.50\textwidth}{!}{\includegraphics{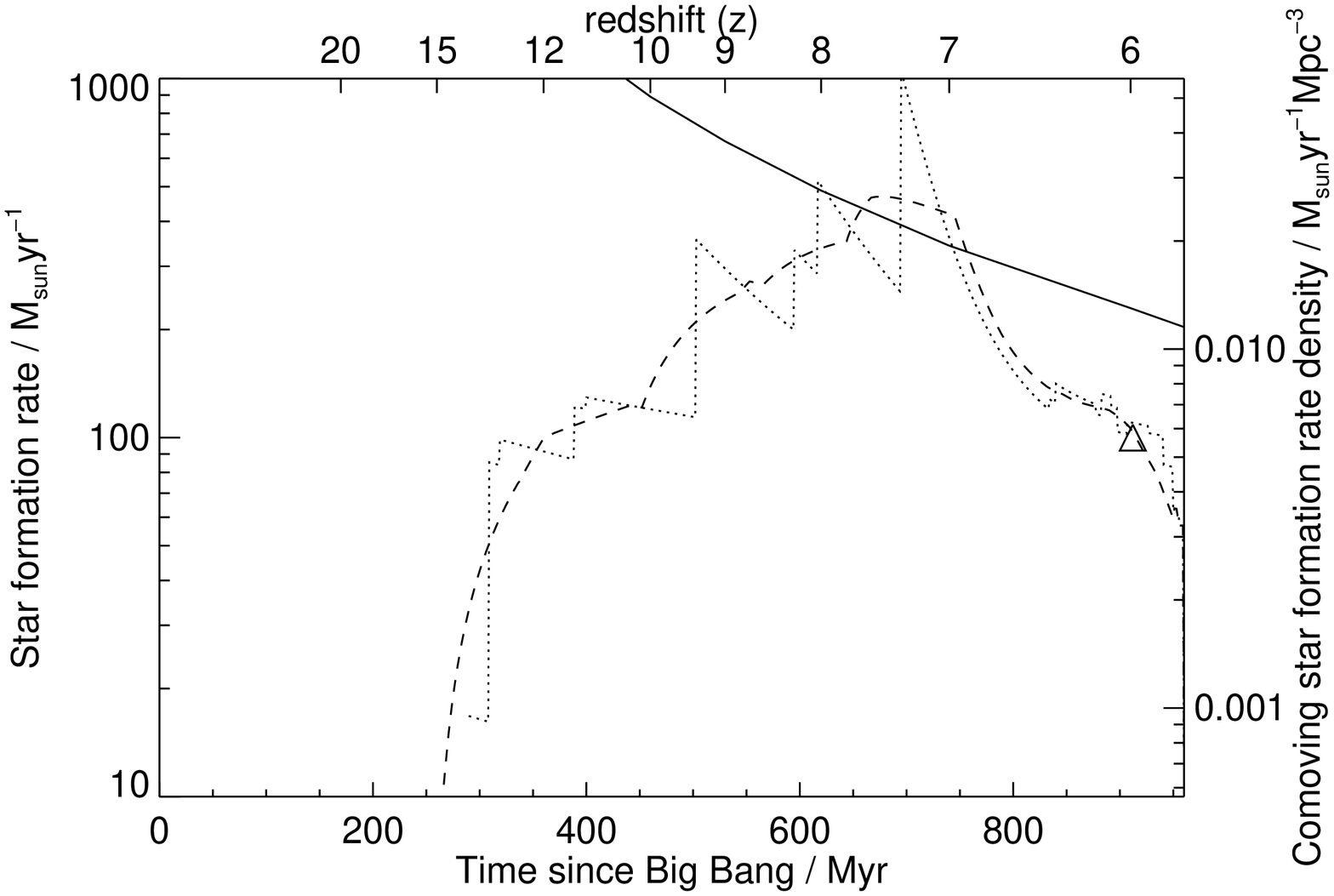}}
\caption{{\bf Left:} The evolution of the stellar mass density -- see
Eyles et al.\ (2007) for details of this compilation and references to the literature. Our measurement from the $i'$-drop galaxies at $z\approx 6$ is marked by
a star. {\bf Right:} The sum of the past
star formation rates for our $i'$-drop sample (dotted curve,
and smoothed over 100\,Myr for dashed curve). The
requirement for reionization is the solid curve (from Madau, Haardt  \& Rees 1999) -- if the escape fraction is high, there is sufficient
UV flux from star formation to achieve reionization at $z\ge 7$ (Eyles et al.\ 2007).}
\label{fig:StellarMassHist}
\end{figure}

\end{document}